# Sécurité alimentaire de l'agriculture indigène guatémaltèque face à l'incertitude sociale et climatique

*Développement participatif d'un modèle couplé d'un système agricole indigène à petite échelle au Tz'olöj Ya' (Guatemala)*

Julien Malard-Adam*

Département de génie des bioressources, Université McGill, Montréal, Québec, Canada   Unité mixte de recherche G-EAU, Institut de recherche pour le développement, Université de Montpellier, Montpellier, France ; Directorat d'enseignement et de vulgarisation, Université Agricole du Tamil Nadu (விரிவாக்க கல்வி இயக்ககம், தமிழ்நாடு வேளாண்மைப் பல்கலைக்கழகம்), Inde,  adresse courriel: julien.malard@mail.mcgill.ca

Jan Adamowski

Département de génie des bioressources, Université McGill, Sainte-Anne-de-Bellevue, Québec, Canada
Adresse courriel : jan.adamowski@mcgill.ca

Héctor Tuy

Instituto de Agricultura, Recursos Naturales y Ambiente (IARNA), Universidad Rafael Landívar, Armita' (Guatemala), Guatemala Centroamérica
Adresse courriel : htuy@fulbrightmail.org

Hugo Melgar-Quiñonez

Institut Margaret A Gilliam pour la Sécurité Alimentaire, Université McGill, Montréal, Québec, Canada
Adresse courriel : hugo.melgar-quinonez@mcgill.ca

**Résumé**

Tandis que les changements climatiques menacent l'agriculture à petite échelle, l'importance des méthodes de modélisation participatives qui peuvent prendre en compte les aspects humains et environnementaux de ces systèmes agricoles prend de l'ampleur. Cette étude présente un modèle socioéconomique des systèmes alimentaires et environnementaux de la région majoritairement indigène du Tz'olöj Ya' au Guatemala, développé de manière participative avec les parties prenantes de la région et ensuite couplé à un modèle externe des cultures. Le modèle est par la suite utilisé afin d'analyser l'impact des changements climatiques futurs sur le système, autant que leur interaction avec différentes politiques proposées par les parties prenantes. L'analyse démontre que les cycles de rétroaction entre composantes environnementales et humaines au sein du modèle des dynamiques des systèmes mènent à des réponses parfois contre-intuitives aux politiques proposées. En un même temps, l'utilisation d'un modèle des cultures externe facilite la représentation, de manière réaliste, des impacts des changements climatiques sur les cultures. Cette analyse est la première à utiliser un modèle socio-économique des dynamiques des systèmes, couplé avec un modèle externe des cultures, pour analyser un système alimentaire socio-environnemental et la sécurité alimentaire qui en dépend.

**Abstract**
Given the increasing pressures exerted by climate change on small-scale agriculture, the importance of participatory modelling methodologies that can consider both the human and environmental components of these systems has become more and more evident. The current study presents a socioeconomic system dynamics model of the food and environmental systems of the predominantly Indigenous region of Tz'olöj Ya', Guatemala. The model was built in a participatory manner with stakeholders from the region and was then coupled to an external crop growth model before being applied to the analysis of the impact of future climate change and its potential interactions with various stakeholder-proposed policies. The analysis identified several feedback loops between environmental and human components of the system that can lead to counterintuitive responses to the proposed policies. At the same time, the use of an external crop growth model allowed for a more realistic, yet still easily implementable, representation of the impacts of climate change on crop production. This analysis is the first to use a socioeconomic system dynamics model coupled with an external crop growth model to analyse food security in the context of a local socio-environmental food system.

**T1 Introduction**

Les systèmes agricoles à petite échelle, de même que les communautés qui en dépendent, se voient de plus en plus menacés par les changements climatiques à l'échelle mondiale. Tandis que les modèles agricoles et climatiques (Jones et Thornton, 2000 ; Brisson et al., 2003 ; Jones et al., 2003 ; de Wit, 2019) peuvent donner des prévisions quant aux changements environnementaux que nous réserve l'avenir, les interactions entre ces changements et les réponses des sociétés qui dé termineront le coût sociétal final de ceux-ci demeurent plus complexes à prédire (Srinivasan et al., 2018 ; Di Baldassarre et al., 2019).

Plusieurs approches de modélisation, dont nombreuses de format participatif (Chlous-Ducharme et Gourmelon, 2012 ; Diallo et al., 2014 ; Srinivasan et al., 2018 ; Bojórquez-Tapia et al., 2019 ; Iwanaga et al., 2020 ; Perrone et al., 2020), visent à améliorer notre compréhension de ces processus. Les systèmes agricoles à petite échelle étant particulièrement complexes et comprenant des relations et de nombreuses rétroactions entre les domaines environnemental, sociétal et économique, de plus en plus de travaux visant à analyser ces systèmes ont recours au cadre conceptuel de la modélisation des dynamiques des systèmes. Cette méthodologie, qui permet de modéliser de manière explicite les variables socio-économiques autant qu'environnementales de même que les rétroactions entre ces variables et les comportements contre-intuitifs engendrés par ces dernières, connaît ainsi une popularité croissante dans le domaine de la modélisation participative des systèmes agricoles. Une revue détaillée de l'utilisation de la modélisation des dynamiques des systèmes afin de représenter les systèmes socio-économiques fut présentée par Elsawah et al. (2017).

De nombreuses études précédentes font également appel à la méthodologie des dynamiques des systèmes pour analyser les relations entre les changements climatiques et la résilience des systèmes socio-environnementaux. Parmi celles-ci, nous recensons les travaux de Chapman et Darby (2016), qui l'ont utilisée pour analyser la durabilité de différentes méthodes de culture du riz au Vietnam, de même que ceux de Stojkovic et Simonovic (2019) qui l'ont appliquée pour prédire la génération d'hydroélectricité d'un barrage en Serbie sous différents scénarios climatiques. Ces études analysent les processus par lesquels les humains interagissent avec les systèmes agricoles et environnementaux dont ils dépendent et les manières dont ceux-ci répondent aux influences humaines. Elles peuvent aussi démontrer comment les sociétés humaines tentent de s'adapter aux changements climatiques et visent à identifier là où les limites sociétales et environnementales à l'adaptation risquent d'être franchies, comme le démontre une analyse de la résilience du système socio-économique estuaire du Bangladesh aux changements économiques et climatiques (Hossain et al., 2017). De manière similaire, Nguyễn Thành Tựu (2018) a utilisé une méthode des dynamiques des systèmes pour prédire les impacts des changements climatiques sur le système de riziculture du delta du Mékong (*đồng bằng sông Cửu Long*) au Vietnam.

Des études précédentes par les auteurs du présent article (Malard et al., 2023a ; 2023b) présentent le développement participatif d'un modèle des dynamiques des systèmes pour les régions de « l'altiplano » (le plateau) guatémaltèque, analyse qui comporte une étude de cas dans la municipalité de Concepción au Tz'olöj Ya' (aussi connue sous le nom de Sololá selon son appellation hispanophone). Ces régions faisant face à des niveaux d'insécurité alimentaire élevés depuis des décennies, elles portent donc un intérêt particulier pour l'application d'une méthodologie axée sur les dynamiques des systèmes, capable de déceler les rétroactions et les délais complexes qui peuvent gêner les initiatives qui manquent d'intégration entre les secteurs socio-économiques et environnementaux.

Les modèles des dynamiques des systèmes présentent cependant d'importantes limitations dans le contexte de la modélisation des systèmes physiques tels que la croissance des cultures ou les changements climatiques (Malard et al., 2017 ; Ammar et Davies, 2019), pour lesquels il existe déjà de nombreux modèles spécialisés. Les études ayant recours à la modélisation des dynamiques des systèmes dans de tels contextes se voient donc en majeure partie limitées à une représentation simpliste du domaine environnemental lorsque celui-ci est implémenté dans le langage de simulation des dynamiques des systèmes. Tandis que de telles représentations peuvent être suffisantes pour une analyse générale des dynamiques, processus et tendances du système (Stojkovic et Simonovic, 2019), elles s'avèrent insuffisantes lorsque l'on désire analyser la réponse du système face à des conditions plus complexes, tel l'impact des changements climatiques sur les cultures. Ces limitations affectent donc le potentiel d'application des modèles des dynamiques des systèmes dans le cadre d'analyses plus détaillées de l'impact des changements climatiques sur les systèmes socio-environnementaux.

En ce sens, le modèle des cultures développé lors d'études précédentes au Guatemala (Malard et al., 2023a ; 2023b) demeure relativement simpliste et n'inclut, selon la méthodologie d'Inam et al. (2017a), que l'impact du déficit d'évapotranspiration annuel afin de calculer le rendement du maïs, culture centrale aux systèmes alimentaires locaux et nationaux. Ces limitations compliquent la possibilité d'une analyse plus approfondie de l'impact du climat sur le système agricole, étant donné que les changements climatiques, de même que leur impact sur les cultures, vont au-delà du simple total de précipitation et incluent divers facteurs tels que les températures extrêmes et la distribution de la précipitation tout au long de la saison de croissance des cultures.

Le couplage entre les modèles des dynamiques des systèmes et les modèles physiques peut ainsi faciliter le développement de meilleurs modèles socio-environnementaux intégrés (Inam et al., 2017a ; 2017b ; Malard et al., 2017). Lors de cette approche, les relations socio-économiques sont représentées dans le modèle (graphique) des dynamiques des systèmes, tandis que les processus physiques et environnementaux se voient relégués à un modèle physique externe, les résultats des deux composantes étant ensuite échangés de manière dynamique entre les modèles couplés lors de la simulation. Tandis qu'une telle approche, comportant un modèle des dynamiques des systèmes couplés à un modèle physique, a été appliquée à plusieurs reprises par le passé (entre autres à la gestion de la salinité des sols au Pakistan (Inam et al., 2017a ; 2017b ; Malard et al., 2017) et à l'intégration des systèmes agropastoraux au Yucatán au Mexique (Patrick Smith et al., 2005 ; Parsons et al., 2011a ; 2011b), aucune application à la modélisation de la malnutrition ou de l'insécurité alimentaire et de leurs liens avec les systèmes agricoles et économiques à petite échelle, ni à la résilience de ces systèmes envers les changements climatiques, n'a été recensée à ce jour.

Cette étude propose donc deux objectifs. Premièrement, l'élaboration d'une méthodologie pour l'utilisation des modèles des dynamiques des systèmes couplés aux modèles agronomiques afin d'analyser la résilience des systèmes agricoles à petite échelle (surtout face aux changements climatiques) et, deuxièmement, l'application de cette méthodologie au développement de politiques résilientes des points de vue agronomique, environnemental et social pour la région du Tz'olöj Ya' au Guatemala.

**T1 Méthodologie**
*T2 Étude de cas*
L'étude présente se déroule dans la municipalité de Concepción, au Tz'olöj Ya' du Guatemala. Cette région est fortement dépendante de l'agriculture (commerciale autant que d'autosuffisance) et recense une très forte présence indigène (99,67% en 2018) (INE, 2019). En outre, les très hautes prévalences d'insécurité alimentaire et de malnutrition chronique chez les enfants à travers les décennies récentes donnent à cette région une importance particulière pour l'analyse systémique des systèmes agroalimentaires et environnementaux.

*T2 Modèle des dynamiques des systèmes*

Le modèle des dynamiques des systèmes a été élaboré selon la méthodologie participative originalement développée et appliquée au Pakistan (Inam et al., 2015 ; Halbe et al., 2018) et ensuite au Guatemala (Malard et al., 2023a ; 2023b). Lors de la première étape du processus participatif, une série d'entrevues individuelles de construction de diagrammes de boucles causales a été effectuée avec la participation de diverses parties prenantes de la région (organisations agricoles, organismes à but non lucratif et représentants gouvernementaux de différents niveaux des départements de sécurité alimentaire, d'agriculture, d'environnement et d'éducation). Les boucles causales ainsi développées par les parties prenantes elles-mêmes ont ensuite été combinées en un seul modèle unifié représentant les points de vue de chaque partie prenante. Ces activités ont été suivies d'un sondage par écrit et d'une série d'ateliers de groupe échelonnés sur une période de six mois qui ont permis le développement continu du modèle et l'ajustement de ses différentes composantes selon les clarifications et les discussions entre les parties prenantes.

[Figure 1]

Le modèle final inclut des modules d'agriculture, d'utilisation du territoire, de population, d'alimentation et d'économie locale. Le modèle agricole modélise le rendement du maïs, culture importante du système dit *awan* (aussi connu selon son nom hispanophone de *milpa*) de co-culture de maïs avec d'autres plantes telles les cucurbitacées et les légumineuses. L'utilisation du territoire inclut trois différentes utilisations des terres (forêt, agriculture et arbustes). Les transformations entre l'une et l'autre sont déterminées selon les politiques de reforestation et les besoins en bois et en production agricole, ce dernier étant à son tour déterminé selon la demande pour la production d'autosuffisance et l'attrait économique relatif de l'agriculture commerciale en comparaison avec celui du travail salarié.

Une description complète de chaque module est présentée dans une étude précédente (Malard et al., 2023b). Un seul changement a été apporté au modèle afin d'effectuer la connexion avec le modèle des cultures externe, soit l'équation déterminant le rendement du maïs. Celle-ci, originalement calculée selon une fonction de perte par stress hydrique, a été modifiée ainsi :

Formule originale : $R = R_0 * F$

Formule modifiée : R = IF THEN ELSE ( R_externe >= 0, R_externe, R0 * F )

Où *R* est le rendement du maïs, *R_externe* la valeur fournie par le modèle externe, *R0* le rendement maximal selon la formule initiale et *F* le facteur de perte par stress hydrique selon la formule initiale. L'initialisation de *R_externe* avec une valeur par défaut négative assure que le modèle des dynamiques des systèmes utilise le rendement du modèle externe uniquement lorsque le modèle est simulé en configuration couplée et utilise la formule initiale lorsqu'il est simulé seul. Dû à la grande diversité (et à la fluctuation saisonnière selon les marchés) des cultures commerciales dans la région et aux difficultés associées à leur représentation individuelle par l'entremise d'un modèle biophysique, le modèle ne considère que le revenu économique généré par superficie d'agriculture commerciale.

*T2 Modèle biophysique*

Le modèle PCSE (de « Python Crop Simulation Environment » en anglais ou « Environnement de modélisation des cultures Python » en français) développé par de Wit en 2019,[1] a été choisi pour implémenter les processus de croissance des cultures du modèle (de Wit et al., 2019). Ce choix découle en majeure partie du fait que PCSE a été implémenté en Python, ce qui facilite de manière importante son intégration avec les outils Python utilisés au cours des analyses de la présente étude. Afin d'effectuer une simulation, PCSE nécessite des données du sol, du climat, de la phénologie de la plante elle-même, et des options de gestion. Pour les sols, les paramètres du sol « EC3-medium fine » de la base de données GGMCI[2] de de Wit ont été empruntés. Suivant la réalité sur le terrain de cette région du Guatemala en ce qui concerne les champs d'*awan*, aucune irrigation n'a été appliquée.

Un défi important à la modélisation des systèmes agricoles à petite échelle est la quantité limitée de jeux de paramètres calibrés disponibles dans la littérature existante pour les variétés des cultures utilisées par les agricultrices et agriculteurs à petite échelle, comparativement à ce qui est disponible pour les variétés utilisées dans les grandes cultures. Ainsi, et suivant une méthodologie précédemment appliquée pour une situation similaire au Mexique (Parsons et al., 2011b), des simulations initiales ont été effectuées avec un jeu de paramètres PCSE disponible pour une variété de maïs tropicale. Trois paramètres du module des cultures de PCSE, notamment TSUM1, TSUM2 et CV0, ont ensuite été ajustés manuellement afin d'assurer que le développement du maïs simulé lors d'une année à conditions climatiques typiques s'apparente au rendement maximal observé et au calendrier de développement tel que rapporté par le ministère de l'Agriculture, de l'Élevage et de l'Alimentation du Guatemala (MAGA)[3] pour la région du pays correspondant à notre étude de cas. La simulation du modèle des cultures a été configurée pour débuter le 1$^{er}$ décembre de la première année de simulation du modèle couplé, le maïs étant semé le 1$^{er}$ mai, suivi par une récolte à maturité ou au plus tard après 240 jours. Les détails concernant le traitement des données climatiques suivront dans la section ci-dessous.

---

[1] Il s'agit de la version 5. 4. 2. du Python Crop Simulation Environment (version 5.4.2) développé par de Wit en 2019. Pour avoir accès au modèle PCSE, suivre le lien suivant [En ligne], URL : https://pcse.readthedocs.io/en/5.4.2/search.html

[2] Pour avoir accès à la base de données GGMCI, suivre le lien suivant, [En ligne], URL : https://github.com/ajwdewit/ggcmi

*T2 Connexion des modèles*

Les simulations ont été implémentées avec le logiciel *Tinamït*, un paquet Python qui facilite la construction et la simulation des modèles des dynamiques des systèmes couplés à un modèle physique (Malard et al., 2017). En comparaison avec les autres méthodes disponibles pour coupler des modèles des dynamiques des systèmes avec des modèles physiques externes, l'approche de *Tinamït* offre l'avantage d'être plus flexible et rapide à implémenter, tandis que les méthodes utilisant des macros en Excel (Inam et al., 2017a ; 2017b) ou des extensions pour Vensim (Patrick Smith et al., 2005 ; Parsons et al., 2011b) ont été utilisées dans le même but par le passé. Ces approches sont généralement peu flexibles et nécessitent la modification du code source (et possiblement une recompilation) à chaque modification d'une variable connectée. Au contraire, *Tinamït* permet une connexion automatisée des variables des deux modèles et se charge ensuite d'échanger leurs valeurs lorsqu'une simulation est amorcée. En outre, ce logiciel permet aussi la connexion automatisée, à travers le paquet de gestion de données climatiques تقدير (taqdir[4]) (Malard, 2019), de variables du modèle physique ou des dynamiques des systèmes avec des bases de donné

---

[3] Pour plus d'informations, voir le site du *Mesoamerican Famine Early Warning System Network* (MFEWS), [En ligne], URL : https://fews.net/sites/default/files/crop_assessment_cards_maize_beans.pdf

Ce système d'alerte précoce de surveillance de la sécurité alimentaire a été établi en 1985 par l'Agence américaine pour le développement international (USAID). Pour plus d'informations, voir le site *du Famine Early Warning Systems Network* (FEWS NET), [En ligne], URL : https://fews.net/

[4] Pour plus d'informations sur le paquet de gestion des données climatiques taqdir, voir le lien suivant, [En ligne], URL : https://taqdir.readthedocs.io/fr/latest/malumat.html

es climatiques externes, y compris des prévisions de changement climatique fournies par MarkSim (Jones et Thornton, 2000).

Le modèle des dynamiques des systèmes a été construit avec le logiciel visuel de modélisation des dynamiques des systèmes Vensim PLE[5], mais toutes les simulations ont été effectuées par l'entremise de *Tinamït* et de son moteur de simulation par défaut, le simulateur de modèles des dynamiques des systèmes en Python PySD (Houghton et Siegel, 2015). Tandis que *Tinamït* peut également simuler des modèles couplés avec l'engin de simulation de Vensim, ces fonctionnalités nécessitent les librairies uniquement disponibles dans la version payante DSS du logiciel Vensim, librairies qui, en tout cas, ne fonctionnent que sur le système d'exploitation Windows. L'expérience des auteurs est aussi que l'outil PySD s'avère généralement nettement plus rapide que l'engin de simulation de Vensim lors d'une simulation couplée.

Pour permettre la connexion du modèle des cultures PCSE et du modèle des dynamiques des systèmes en Vensim avec *Tinamït*, une « enveloppe » spécifique à l'utilisation du modèle PCSE dans cette étude a été développée selon la spécification de *Tinamït*. Cette enveloppe gère la création du modèle PCSE pour une région donnée du Guatemala et effectue les connexions nécessaires entre *Tinamït* et PCSE pour le transfert des données des variables climatiques et du rendement modélisé chaque année. Le Tableau 1 ci-dessous résume les variables échangées entre les différentes composantes du modèle couplé.

[Tableau 1]

*T2 Données et changements climatiques*

Les données climatiques historiques ont été obtenues depuis la base de données d'observations satellitaires *NASA Prediction of Worldwide Energy Resources* (NASA POWER)[6], tandis que les prévisions climatiques ont été obtenues du Projet de comparaison de modèles couplés (PCMC phase 5) du Programme mondial de recherche sur le climat (PMRC) par l'entremise du logiciel MarkSim (Jones et Thornton, 2000). Ces sources de données étant toutes deux incluses en tant que sources par défaut dans le logiciel تقدير (taqdir), l'extraction et l'intégration de leurs données dans le modèle étaient possibles en quelques lignes de code Python.

---

[5] Version 7. 3. 5. de *Vensim PLE*, 2017, Ventana Systems.
[6] Pour plus d'informations, voir le site du *POWER Project* de la NASA, [En ligne], URL : https://power.larc.nasa.gov/

Un défi important lors de l'application de prévisions climatiques des modèles globaux au niveau local concerne la correction du biais intrinsèque des prévisions à grande échelle comparées aux observations locales. En somme, la moyenne autant que la variabilité des prévisions générées avec un scénario sans changements climatiques devaient s'apparenter à celles des observations historiques correspondantes pour la région. Lors de cette étude, les prévisions de température (maximale, minimale et moyenne quotidiennes) et de précipitation ont été ajustées en utilisant l'année 2010 comme date limite (inclusive) pour les données historiques.

Pour la température, la méthode de correction linéaire telle que présentée par Luo et al. (2018) a été appliquée. Lors de cette correction, la température quotidienne corrigée $T_{m,j}^{corr}$ pour le jour $j$ du mois $m$ est donnée par l'équation suivante :

$$T_{m,j}^{corr} = T_{m,j} + [\bar{T}_{obs,m} - \bar{T}_{mod,m}]$$

Où $T_{m,j}$ est la température prédite pour le jour $j$ du mois $m$ sous un certain régime de changement climatique, $\bar{T}_{obs,m}$ est la moyenne de la température historique observée pour le mois $m$ et $\bar{T}_{mod,m}$ la moyenne de la température prédite par le modèle lorsqu'il est simulé sans changements climatiques (climat historique). La même approche fut utilisée pour l'évapotranspiration de référence.

La situation se complique légèrement pour le cas de la précipitation. Cette variable représente une combinaison de deux composantes primaires, soit, premièrement, s'il y a eu de la pluie au cours d'une journée donnée (valeur binaire), et secondement, le cas échéant, la quantité de pluie (valeur positive). Les méthodes recensées par Luo et al. (2018) ne s'appliquent que lorsque les prévisions du modèle incluent un trop grand nombre de journées pluvieuses en comparaison avec les observations historiques, tandis que le contraire est observé dans le cas de l'étude de Vrac et al., (2016). Dans un tel cas, une méthode différente nommée « Élimination stochastique de singularités », où chaque journée sèche se voit assignée d'une valeur aléatoire entre zéro et un seuil critique représentant le minimum de la pluie quotidienne recensée dans les données, est plus appropriée (Vrac et al., 2016). Des méthodes traditionnelles de correction de biais peuvent ensuite être appliquées à l'ensemble des données, à la suite desquelles les jours qui ont toujours des valeurs inférieures au seuil critique se voient réattribués d'une précipitation nulle. Cette approche, combinée avec la méthode de correction linéaire suivie d'une variante de la transformation par puissance (Luo et al., 2018), a été utilisée dans cette étude.

La première correction linéaire ajuste la moyenne de précipitation mensuelle :

$$P_{m,j}^{corr} = P_{m,j} * \left[\frac{\bar{P}_{obs,m}}{\bar{P}_{mod,m}}\right]$$

Où $P_{m,j}$ est la précipitation modélisée pour le jour $j$ du mois $m$, $\bar{P}_{obs,m}$ la moyenne du total de la précipitation historique observée pour le mois $m$ et $\bar{P}_{mod,m}$ la moyenne du total de la précipitation modélisée pour le même mois.

Ensuite, une transformation de puissance a été appliquée pour corriger la variance de la précipitation. Contrairement à la description de cette transformation par Luo et al. (2018), la correction a été appliquée afin de minimiser la différence entre les valeurs observées et modélisées de la précipitation annuelle cumulative de chaque mois de l'année. Cette décision a été prise suite à une analyse initiale qui démontrait qu'une simple correction au niveau de chaque mois individuel se soldait fréquemment en un mélange de mois secs et humides au cours de la même année et ne représentait donc pas de manière satisfaisante la variabilité interannuelle en précipitation, cause clef des sécheresses et des pertes des récoltes dans la région.

Selon cette approche, une mesure $s_m$ de la différence entre la variabilité interannuelle observée et celle modélisée après correction est établie pour chaque mois :

$$s_m = \sigma\left(\sum_{i=0}^{m} P_{obs,i}\right) - \sigma\left[\left(P_{mod,m}\right)^{b_m} * \frac{\bar{P}_{mod,m}}{\left(\bar{P}_{mod,m}\right)^{b_m}} + \sum_{i=0}^{m-1} P_{mod,i}\right]$$

Où σ indique l'écart type interannuel, $P_{obs,i}$ est l'observation de précipitation pour le mois $i$, $P_{mod,i}$ représente la précipitation modélisée déjà corrigée pour le mois $i$, $P_{mod,m}$ la précipitation du mois $m$, $b_m$ le facteur de correction pour le mois $m$ et $\bar{P}_{mod,m}$ la moyenne interannuelle du total de la précipitation modélisée pour le mois $m$.

L'algorithme tente ensuite d'optimiser $b_m$ afin de minimiser la valeur absolue de $s_m$. La précipitation corrigée pour chaque jour $j$ du mois $m$ est ensuite calculée selon la formule suivante :

$$P_{m,j}^{corr} = P_{m,j} * \left[\frac{\left(P_{mod,m}\right)^{b_m}}{P_{mod,m}}\right] * \left[\frac{\bar{P}_{mod,m}}{\left(\bar{P}_{mod,m}\right)^{b_m}}\right]$$

Où $P_{m,j}$ est la précipitation modélisée pour le jour $j$ du mois $m$ et les autres variables sont définies comme ci-dessus.

En commençant par le premier mois de l'année (janvier), la procédure est répétée pour chacun des 12 mois afin d'ajuster les valeurs de précipitation de chaque mois de l'année. Le mois de janvier a servi de premier mois de l'année, car il se situe dans une période sèche et sans culture d'*awan* au Tz' olöj Ya' et offre par conséquent une bonne référence de départ pour la précipitation annuelle d'un point de vue agronomique.

Tandis que les valeurs d'évaporation et de transpiration, de même que de la pression de vapeur, sont incluses dans la base de données d'observations POWER, elles ne font pas partie des prévisions générées par MarkSim. Les valeurs d'évaporation et de transpiration pour les prévisions climatiques ont donc été calculées selon l'équation Penman-Monteith telle qu'implémentée dans PCSE, tandis que la pression de vapeur a été calculée sur la base de la température minimum journalière selon l'implémentation en Python dans le logiciel agroécologique *Tiko'n*[7] (Malard et al., 2020) de la méthode initialement implémentée en FORTRAN par le modèle de croissance des cultures DSSAT (Jones et al., 2003).

## *T2 Analyses de scénario*

Six différents scénarios de politiques ont été appliqués en combinaison avec cinq scénarios de changement climatique (aucun changement, Trajectoire représentative de concentration (TRC) 2,6 ; TRC 4,5 ; TRC 6,0 et TRC 8,5), pour un total de 35 différentes combinaisons, y compris le scénario de base. Les TRCs représentent différents scénarios d'émissions de gaz à effet de serre selon le cinquième rapport du Groupe d'experts intergouvernemental sur l'évolution du climat (GIEC), du moins élevé (2,6) au plus élevé (8,5). Les politiques, dont la majorité avait été proposée par les parties prenantes lors des ateliers de groupe (incluant la reforestation, l'autosuffisance pour les intrants agricoles, les achats de groupe, les jardins familiaux et l'éducation), ont été choisies pour représenter une grande gamme d'actions possibles sur les plans économique, environnemental et social. Au cours de ces activités de groupe, les parties prenantes ont discuté du modèle que nous avions co-développé afin d'améliorer la représentation du système agricole dans le modèle et ont recherché des idées de politiques publiques pour améliorer la sécurité alimentaire. Certaines politiques, telles que les jardins familiaux et l'autosuffisance pour les intrants agricoles, étaient déjà en cours de mise en œuvre (quoiqu'à très petite échelle) ; les autres propositions représentent de nouvelles idées qui ont été proposées en tant qu'actions futures potentielles. Tableau 2 ci-dessous présente un résumé de l'implémentation de chacune de ces politiques.

---

[7] Le logiciel agroécologique *Tiko'n* permet de développer des modèles de réseaux agroécologiques (interactions trophiques dans un système agricole). Ces modèles de réseaux agroécologiques comprennent les insectes herbivores, les cultures et les insectes carnivores, ainsi que les actions de gestion humaine (biocontrôle, irrigation, application d'engrais, etc.). Pour avoir accès à la bibliothèque *Tiko'n*, suivre le lien suivant, [En ligne], URL : https://github.com/julienmalard/Tikon

[Tableau 2]

*T2 Outils informatiques*

Le modèle et toutes les analyses ont été implémentés en Python 3,7 à l'aide des librairies scipy (Pauli et al., 2020), statsmodels (Seabold et al., 2010), PCSE (de Wit, 2019), xarray (Hoyer et al., 2017), numpy (Harris et al., 2020), matplotlib (Hunter, 2007) et tinamit (Malard et al., 2017). Le modèle, les données et tout code informatique d'analyse sont disponibles en tant que matériel supplémentaire (doi : 10.5281/zenodo.10221534).

**T1 Résultats**

Les résultats de l'étude démontrent une différence marquée entre les conclusions obtenues avec un modèle des dynamiques des systèmes utilisé en isolation et celles obtenues à l'aide du modèle couplé incluant un modèle de croissance des cultures. L'analyse des interactions entre scénarios de changements climatiques et politiques publiques révèle d'importantes différences entre l'efficacité, mais aussi la résilience de différentes options, et suggère que les actions socio-économiques qui visent à s'adresser à la cause d'un enjeu environnemental peuvent être plus efficaces que des politiques s'adressant plutôt à ses symptômes.

*T2 Utilité d'une approche couplée*

L'utilisation d'un modèle couplé pour l'analyse a permis d'incorporer une représentation plus réaliste des systèmes agricoles qu'avec un modèle des dynamiques des systèmes seul. La Figure 2 montre le risque de perte des cultures au fil des ans selon le scénario de changements climatiques et le modèle utilisé : une valeur de 1 indiquerait une perte garantie de la culture, tandis qu'une valeur de 0 n'indiquerait aucun risque de perte de la culture.

[Figure 2]

Comme démontré par la Figure 2, le modèle couplé prédit en moyenne moins de pertes de cultures, mais distingue mieux entre les différents scénarios climatiques. Cette différence découle probablement de l'inclusion de la variabilité intra-annuelle de la pluie dans le modèle couplé. Les déficits de précipitation encourus pendant la saison sèche affectent le bilan d'eau total, mais n'ont pas le même impact qu'une sécheresse en période critique pour la culture du maïs. Ces nuances potentielles des impacts des changements climatiques sur le système socio-économique n'auraient pu être prises en compte sans l'utilisation d'un modèle couplé. En outre, le couplage des modèles permet de considérer cette variabilité intra-annuelle dans la simulation même si le modèle des dynamiques des systèmes lui-même n'a qu'une résolution temporelle annuelle.

La Figure 3 ci-dessous démontre l'influence du modèle des cultures sur la partie socio-économique du modèle, en l'occurrence, sur la progression de la croissance de la population (gauche : utilisation du modèle des dynamiques des systèmes seul ; droite : utilisation du modèle couplé). Sans le modèle des cultures, les plus grandes pertes de rendement prédites par l'implémentation du processus de croissance des cultures dans le modèle des dynamiques des systèmes se traduisent en un taux de migration plus important et, par conséquent, une croissance populationnelle moins forte que lorsque le modèle couplé est utilisé.

[Figure 3]

Cette approche ouvre aussi la porte à des analyses plus détaillées (intrants agricoles, différentes cultures), limitées uniquement par les capacités du modèle des cultures utilisé et les données régionales disponibles. D'une part, de telles analyses, tout comme celle réalisée dans cette recherche, n'auraient pas été possibles avec l'utilisation d'un modèle des dynamiques des systèmes seul. D'autre part, une analyse basée uniquement sur un modèle des cultures ne saurait prendre en compte les composantes socio-économiques du système et les réponses complexes de la société face aux changements climatiques et aux autres défis.

*T2 Changements climatiques et résilience*

L'utilisation d'un modèle couplé a également permis de comparer les comportements potentiels du système socio-environnemental sous des combinaisons de différents scénarios climatiques et de politiques publiques.

[Figure 4]

La Figure 4 ci-dessus présente l'impact de différentes combinaisons de politiques et de scénarios de changements climatiques auprès de plusieurs variables d'intérêt environnemental, économique ou social.

[Figure 5]

Seules les politiques de salaire minimum ou d'éducation universelle mènent à d'importants changements dans les variables d'intérêt. Elles sont même plus efficaces que la politique de reforestation envers l'augmentation de la couverture forestière. Une analyse temporelle (voir la Figure 5 ci-dessus) démontre que tandis que l'implémentation d'un meilleur salaire minimum s'avère plus efficace au long terme, la reforestation offre néanmoins des résultats plus rapides, quoique moins durables. Ces résultats suggèrent donc la possibilité de combiner des politiques pour maximiser les impacts autant au court terme que dans le futur plus lointain. Il est toutefois important de préciser que les politiques d'éducation et de salaire minimum entraînent toutes les deux une forte migration hors de la région d'étude, plus fort même que l'impact graduel que des changements climatiques de plus en plus sévères occasionnent sur les taux de migration. La Figure 6 ci-dessous, qui résume deux boucles de rétroaction présentes dans les modèles développés de manière participative, éclaircit les processus et rétroactions responsables de ces comportements. Lorsque l'éducation s'améliore, une plus haute demande pour des emplois de qualité mène à une réduction de la pauvreté (cycle positif auto-renforçant), et, si la demande est trop importante pour l'économie locale, à un exode des villages vers les centres urbains (cycle d'équilibre).

[Figure 6]

**T1 Discussion**

De manière générale, les résultats suggèrent que les problèmes environnementaux et sociétaux sont les conséquences des systèmes socio-environnementaux dans lesquels ils émergent et que les approches plus holistiques à la gestion environnementale ont de meilleures chances d'avoir des résultats positifs que des interventions axées uniquement sur les symptômes.

La participation a également été une partie clef de la génération des résultats de même que de l'acceptation des conclusions de la recherche par les parties prenantes elles-mêmes. Les politiques publiques analysées dans cette étude, dont la reforestation, l'autosuffisance des intrants agricoles, les achats de groupe d'intrants agricoles, et les jardins familiaux proviennent des recommandations des parties prenantes que celles-ci ont énoncées au cours des ateliers de modélisation participative. Ceci a transformé l'activité de modélisation non seulement en exercice d'intérêt théorique ou scientifique, mais aussi en activité d'intérêt réel pour les participants, car les résultats ont permis d'évaluer le potentiel de succès des différentes actions qu'ils envisageaient d'entreprendre ou bien qu'ils entreprenaient déjà sur le territoire. Ces dynamiques sont devenues apparentes dès la première activité de modélisation de groupe, lorsque l'importance portée sur « l'éducation » et la « sensibilisation » citoyenne, discours plutôt axé sur une vision individualiste envers les solutions aux problèmes sociétaux, a été remise en question par les participants eux-mêmes en faveur de politiques publiques s'attaquant aux enjeux environnementaux plus systémiques, telle la reforestation.

Il est intéressant de noter que la grande majorité des politiques proposées par les parties prenantes n'offrent que des petites améliorations à la situation de sécurité alimentaire dans la région d'étude. Au contraire, ce sont les politiques les plus fondamentales (mais aussi plus difficilement envisageable au niveau de la gouvernance locale) telles que de meilleurs systèmes éducatifs et conditions de travail qui se montrent les plus efficaces et résilientes aux changements climatiques. Ces résultats soulignent l'importance de politiques publiques fortes et coordonnées à l'échelle nationale afin de lutter contre l'insécurité alimentaire. Les résultats de l'article soulignent aussi l'importance, dans le cadre d'une telle analyse socio-environnementale, d'inclure une large gamme de variables d'intérêt, car les mêmes changements auprès des mêmes variables peuvent être plus ou moins désirables selon le contexte. Une diminution dans la population, par exemple, peut être le produit d'une politique robuste d'anti-pauvreté, d'une diminution du taux de fertilité, mais également le résultat d'une migration forcée catastrophique.

De manière similaire, une diminution du taux de pauvreté représente un développement positif si elle est due à une amélioration des conditions de vie et des salaires des résidents, mais pourrait être moins positive si elle reflète un exode populationnel suivant une perte agricole, ne laissant que les résidents les mieux nantis derrière. Ce constat tient également dans le cas des dynamiques forestières : une augmentation de la couverture est désirable si elle représente une meilleure prise en charge environnementale dans la région, mais n'est pas positive si elle est le résultat d'un abandon total de la région suivant une catastrophe environnementale ou autre, tel que cela a déjà été le cas au nord du Guatemala il y a plus d'un millénaire (Gill et al., 2007 ; Evans et al., 2018). Une analyse des dynamiques des systèmes encourage les parties prenantes autant que les chercheurs et chercheuses externes à réfléchir explicitement à chacun de ces processus et rétroactions parfois moins intuitifs et, par conséquent, à développer des politiques plus intégrées qui tiennent davantage compte des réalités socio-environnementales complexes de la région d'étude.

La qualité des données demeure toujours un enjeu dans le cadre d'études portant sur les systèmes socio-environnementaux où des données temporelles précises correspondant à la majorité des variables clefs ne sont pas disponibles. Le manque de données temporelles cause surtout des difficultés au moment de la calibration et la validation des modèles des dynamiques des systèmes, dont la force est justement la génération de prévisions des dynamiques de ces systèmes aux moyen et long termes. L'étude précédente ayant mené au développement du modèle des dynamiques des systèmes utilisé dans la présente étude (Malard et al., 2023b) propose ainsi diverses méthodes de calibration et de construction de confiance dans les résultats des modèles des dynamiques des systèmes lorsque des données temporelles de haute qualité ne sont pas disponibles, en substituant la variabilité spatiale (plus facilement disponible par l'entremise de sondages nationaux) des données pour la variabilité temporelle. Cette approche de calibration par inférence bayésienne permet d'estimer les valeurs des paramètres socio-économiques du modèle, telles que l'influence de l'éducation sur les revenus familiaux, selon les différences observées entre différentes régions du pays (Malard et al., 2023b).

L'utilisation d'un modèle couplé, quant à elle, permet d'incorporer une représentation plus réaliste des processus environnementaux – en l'occurrence, la croissance des cultures – et ainsi d'obtenir des résultats plus fiables qu'avec un modèle des dynamiques des systèmes seul, qui serait contraint à utiliser une représentation très approximative de ces processus biophysiques. Ainsi, le couplage des modèles permet d'incorporer à la simulation les représentations des processus environnementaux tels qu'ils ont déjà été implémentés dans des modèles biophysiques bien vérifiés et établis de longue date, ce qui évite de devoir les copier (ou bien de les approximer) dans un logiciel de modélisation des dynamiques des systèmes qui n'a pas été conçu pour la modélisation numérique plus exigeante des systèmes biophysiques. Dans la présente étude, l'utilisation d'un modèle externe des cultures a permis d'incorporer les aspects plus détaillés des relations entre climat et rendements (température, humidité des sols, nutriments) et laisse également la porte ouverte à l'ajout d'un modèle des cultures plus sophistiqué ou bien d'options de gestion agricole plus poussées dans le futur. Ces résultats seraient difficilement réalisables sans modèle couplé.

La présente analyse est la première à appliquer une méthodologie de modélisation des dynamiques des systèmes couplée à un modèle physique dans le cadre de l'analyse de la sécurité alimentaire et des systèmes socio-économiques et environnementaux dont celle-ci dépend. Par le biais d'une telle analyse intégrée, il a été démontré que les politiques appliquées au système peuvent parfois générer des impacts très forts sur son comportement, y compris sur des variables très éloignées de celles sur lesquelles elles exercent leur première influence. Une analyse qui n'incorpore pas les rétroactions entre différentes composantes du système socio-environnemental ne saurait déceler de tels impacts. L'approche couplée permet aussi de générer des résultats plus précis que ceux qui pourraient être obtenus à partir d'un modèle des dynamiques des systèmes seul, tout en incorporant les rétroactions entre variables sociales, économiques et environnementales.

**T1 Conclusion**

La modélisation des dynamiques des systèmes est de plus en plus utilisée pour représenter les interactions entre humains et environnement. Dans sa forme participative, elle permet aussi aux parties prenantes de présenter et d'inclure leurs points de vue dans le développement du modèle et dans l'analyse des politiques. Cependant, le langage de modélisation des dynamiques des systèmes reste mal adapté à la représentation des composantes physiques des systèmes environnementaux, telles que la croissance des cultures ou les changements climatiques. La présente étude utilise un modèle participatif des dynamiques des systèmes de la sécurité alimentaire et du système agroalimentaire sous-jacent de la région majoritairement indigène de Concepción, Tz'olöj Ya' (Guatemala). Ce modèle a été couplé à un modèle externe des cultures et utilisé pour simuler l'impact des changements climatiques et de diverses propositions de politiques pour lutter contre la malnutrition. Les résultats témoignent de l'aptitude du modèle des dynamiques des systèmes pour la modélisation des liens complexes entre les domaines environnemental, social et économique, tandis que le modèle des cultures se distingue en apportant des prévisions plus réalistes de la réponse agricole aux changements climatiques.

**Figures**

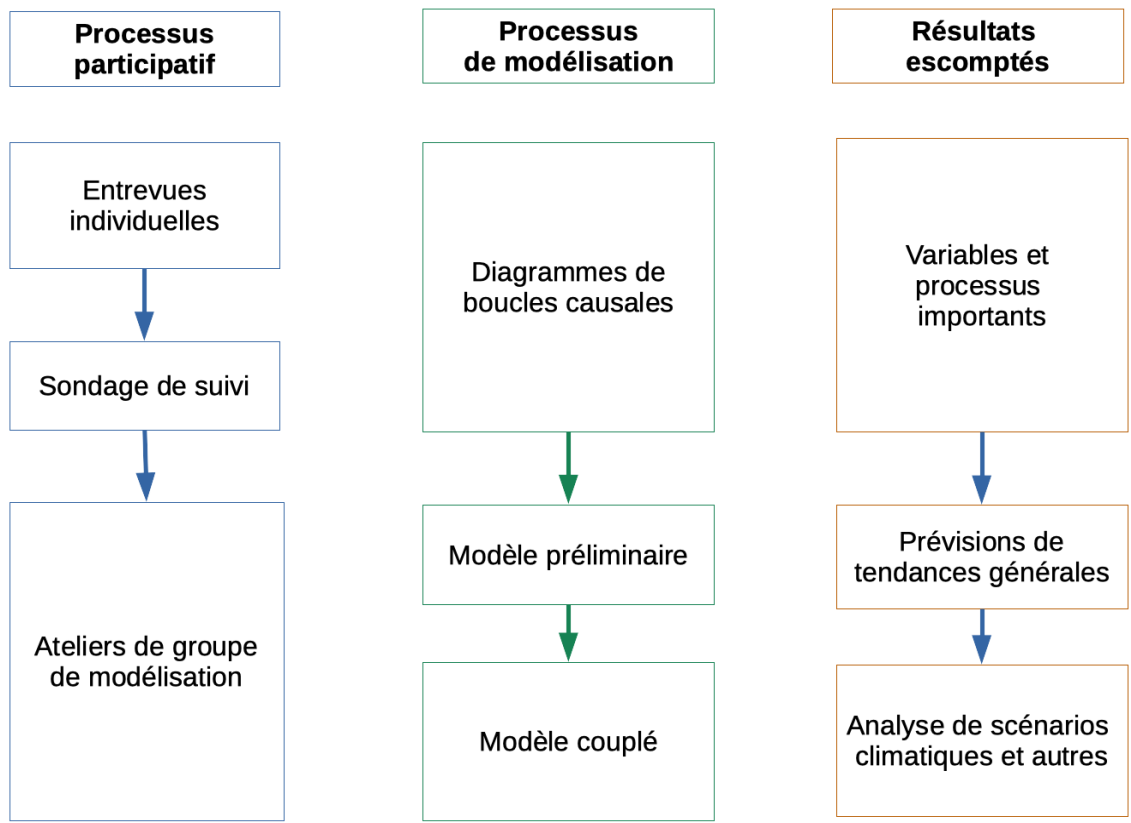

*Figure 1. Diagramme du processus participatif*

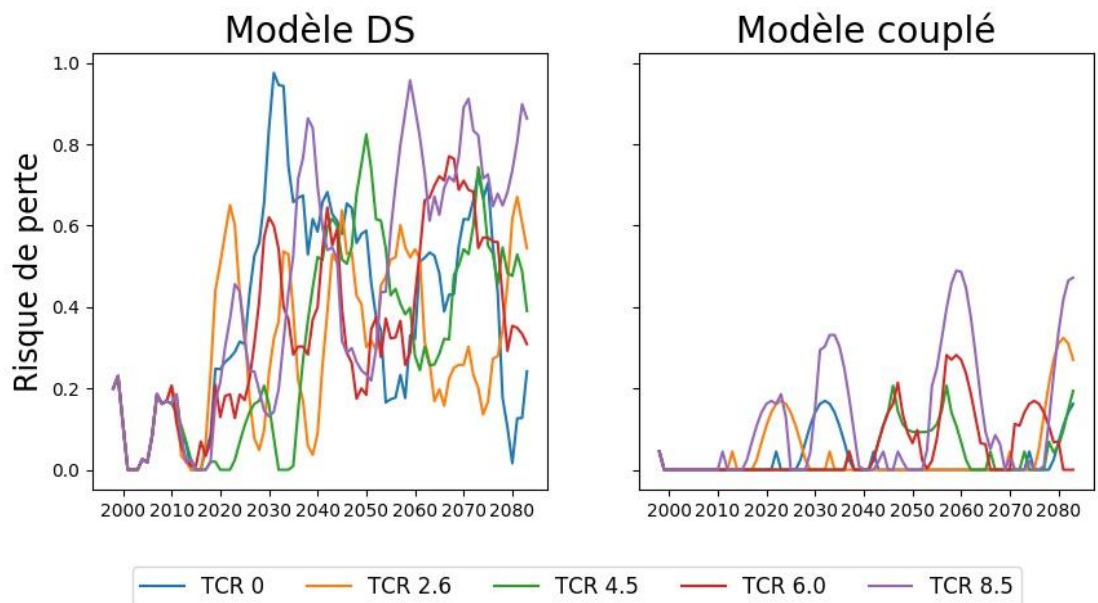

Figure 2. Risque de perte des cultures (lissé) selon l'année de simulation et le scénario de changements climatiques (TRC = Trajectoire de Concentration Représentatif) avec le modèle des dynamiques des systèmes seul (gauche) et couplé (droite).

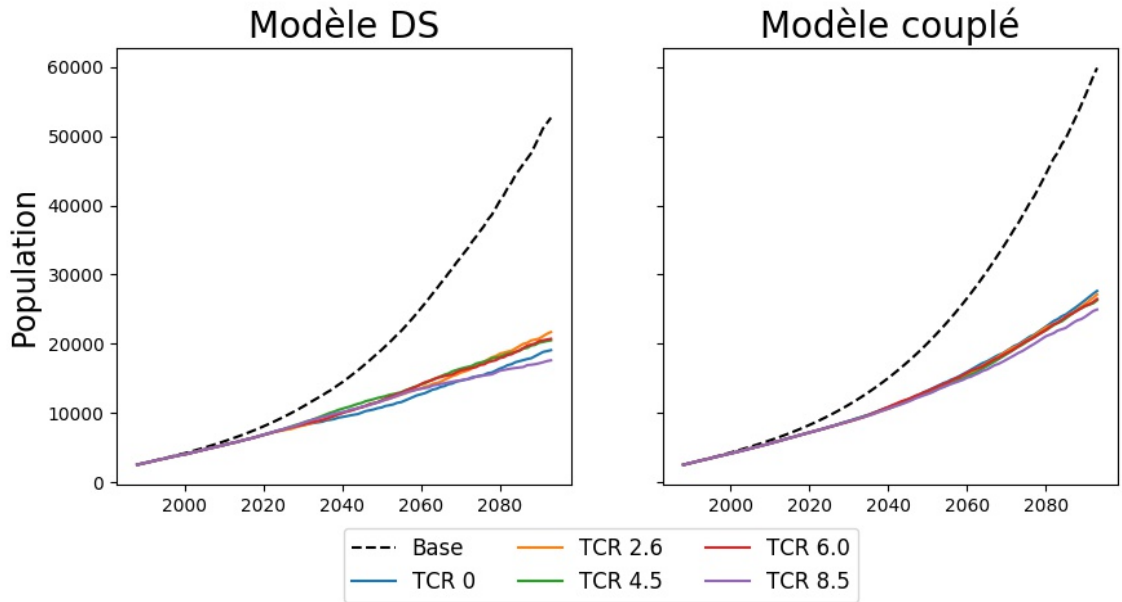

*Figure 3. Influence du modèle des cultures externe sur les projections de population avec la politique d'éducation. À gauche : modèle des dynamiques des systèmes seul ; à droite : modèle couplé. La ligne noire pointillée indique le scénario de base (sans politique ni changements climatiques). TRC = Trajectoire de concentration représentative selon les prévisions du GIEC.*

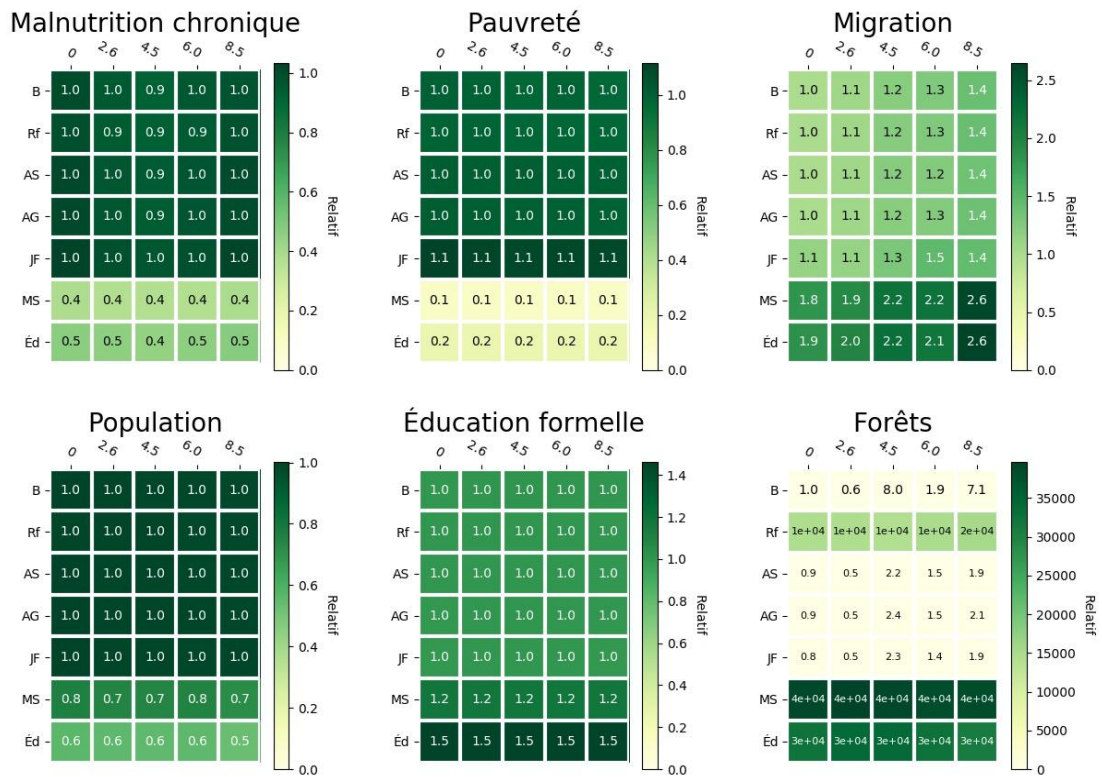

*Figure 4. Impact relatif de chaque combinaison de politique et scénario de changements climatiques sur différentes variables clefs (au cours des 70 dernières années de la simulation). Variables d'intérêt : Malnutrition chronique (chez les moins de 5 ans), pauvreté (moyenne de la différence entre les revenus familiaux et le seuil de pauvreté, uniquement pour les familles en-deçà du seuil), migration (émigration permanente hors de la municipalité), population (de la municipalité), éducation formelle et forêts (couverture forestière de la municipalité). Abréviations de politiques : B=Base, Rf=Reforestation, AS=Autosuffisance produits agricoles, AG=Achats de groupe, JF=Jardins familiaux, MS=Meilleurs salaires, Éd=Éducation. Les chiffres (0 ; 2,6 ; 4,5 ; 6,0 et 8,5) indiquent le scénario TCR (Trajectoire de concentration représentative). Les valeurs présentées sont relatives à la valeur de la même variable pour le scénario B0 (aucune politique publique, aucun changement climatique). Une valeur de 1 indique donc aucun changement entre un scénario donné et le B0 ; une valeur inférieure à 1 indique que la variable en question avait une valeur moindre dans le scénario donnée, et une valeur supérieure à 1 indique que la valeur de la variable était supérieure à celle du scénario B0. À titre d'exemple, la valeur 1,9 à l'intersection de la*

*politique de meilleurs salaires et du TCR 2,6 pour la variable migration indique que la migration, dans ce scénario, à atteint 190% du niveau de migration observé dans le scénario de base B0.*

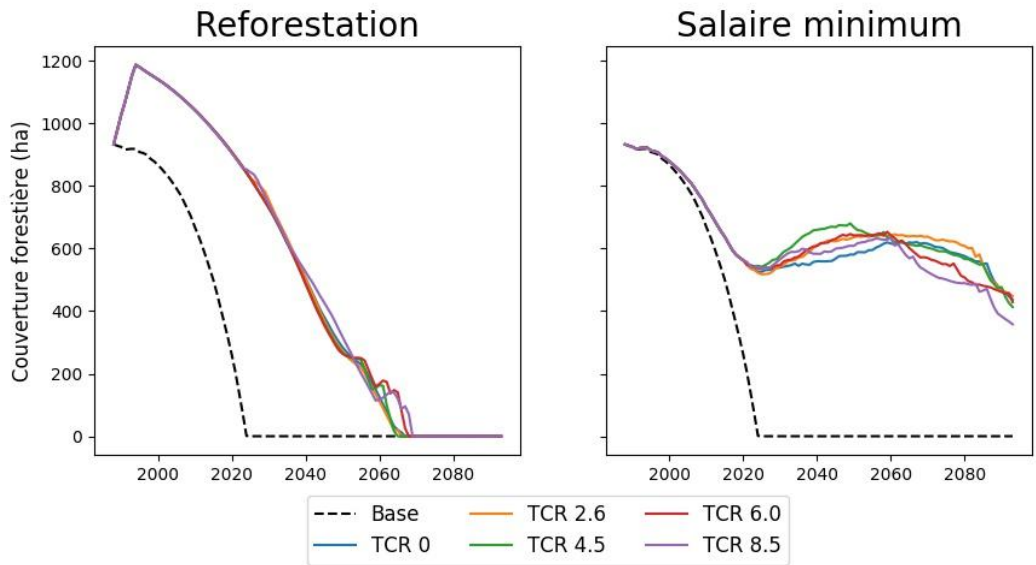

*Figure 5. Effet des politiques de reforestation (gauche) et de salaire minimum (droite) sur la couverture forestière. TCR : Trajectoire de concentration représentative selon les prévisions du GIEC.*

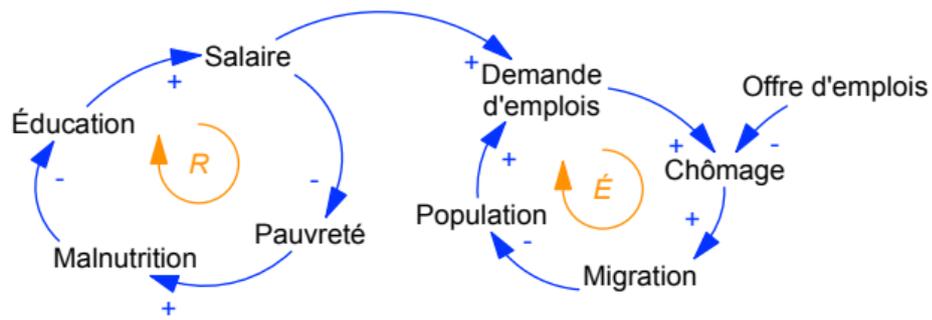

*Figure 6. Diagramme simplifié des dynamiques d'emplois et de salaire dans le modèle. R indique un cycle de rétroaction renforçant, tandis que É indique un cycle d'équilibre.*

**Tableaux**

*Tableau 1. Variables échangées entre les différentes composantes du modèle couplé. PCSE indique le modèle des cultures et MDS le modèle des dynamiques des systèmes. Tinamït/تقدير indique des variables dont les valeurs sont fournies, à travers Tinamït, par le logiciel de gestion des données climatiques تقدير [taqdir].*

| Variable | Code | Unité | Source | Utilisation |
|---|---|---|---|---|
| Rendement du maïs | Rendimiento | kg/ha | PCSE | MDS |
| Évaporation de l'eau | E0 | cm | Tinamït/تقدير | PCSE |
| Évaporation du sol | ES0 | cm | Tinamït/تقدير | PCSE |
| Évapotranspiration des cultures | ET0 | cm | Tinamït/تقدير | MDS, PCSE |
| Pression de vapeur | vap | kPa | Tinamït/تقدير | PCSE |
| Précipitation | precip | mm | Tinamït/تقدير | MDS, PCSE |
| Température moyenne quotidienne | temp_prom | C | Tinamït/تقدير | PCSE |
| Température minimale quotidienne | temp_mín | C | Tinamït/تقدير | PCSE |
| Température maximale quotidienne | temp_máx | C | Tinamït/تقدير | PCSE |

*Tableau 2. Spécification des politiques appliquées au modèle. Q indique quetzales.*

| Politique | Description |
|---|---|
| Reforestation | Reforestation de 50 ha par année |
| Autosuffisance intrants agricoles | Diminution des dépenses agricoles de 75% |
| Achats de groupe d'intrants agricoles | Diminution des dépenses agricoles de 15% |
| Jardins familiaux | Cible d'autosuffisance de 90% |
| Meilleurs salaires | Salaire minimum de 20 Q de l'heure |
| Éducation | Progrès scolaire parfait |